\documentclass[aps,a4paper,twocolumn,10pt,prx]{revtex4-2}

\usepackage{graphicx}
\usepackage{dcolumn}
\usepackage{bm}
\usepackage[utf8]{inputenc}
\usepackage[T1]{fontenc}
\usepackage[english]{babel}
\usepackage{amsmath,amssymb}
\usepackage{textcomp}
\usepackage[toc,page]{appendix}
\usepackage{color}
\usepackage{indentfirst}
\usepackage[separate-uncertainty=true]{siunitx}[=v2]
\usepackage[dvipsnames]{xcolor}
\usepackage{color}
\usepackage[colorlinks=true,allcolors=blue]{hyperref}
\usepackage[normalem]{ulem}
\usepackage{soul}
\usepackage{setspace}
\usepackage{array}
\usepackage{booktabs}
\usepackage{tabulary}
\usepackage{grffile}
\usepackage{float}
\usepackage{lmodern}
\DeclareSIUnit[number-unit-product = {}]{\bact}{bact}
\DeclareSIUnit[number-unit-product = {}]{\pixel}{pixel}

\begin{document}

\preprint{APS/123-QED}

\title{Ostwald-like ripening in the two-dimensional clustering of passive particles \\induced by swimming bacteria}

\author{J. Bouvard}
\author{F. Moisy}
\author{H. Auradou}
\email{harold.auradou@universite-paris-saclay.fr}
\affiliation{Universit\'{e} Paris-Saclay, CNRS, FAST, 91405, Orsay, France.}

\date{\today} 

\begin{abstract}

Clustering passive particles by active agents is a promising route for fabrication of colloidal structures. Here, we report the dynamic clustering of micrometric beads in a suspension of motile bacteria. We characterize the coarsening dynamics for various bead sizes, surface fractions and bacterial concentrations. We show that the time scale $\tau$ for the onset of clustering is governed by the time of first encounter of diffusing beads. At large time ($t \gg \tau$), we observe a robust cluster growth as $t^{1/3}$, similar to the Ostwald ripening mechanism. From bead tracking measurements, we extract the short-range bacteria-induced attractive force at the origin of this clustering.

\end{abstract}

\maketitle

\section{Introduction}
Phase separation and formation of dense domains of self propelled particles are fascinating phenomena in active matter~\cite{theurkauff_dynamic_2012,palacci_living_2013,redner_structure_2013,buttinoni_dynamical_2013,wysocki_cooperative_2014,stenhammar_phase_2014,stenhammar_activity-induced_2015,cates_motility-induced_2015,Bialke2015,Ginot2018,driscoll_leveraging_2019,Paoluzzi2022}.
The structural and dynamical properties of the clusters depend on the particle density, their motility characterized by the persistence time of their trajectories and the existence of attractive mutual interactions. When passive particles are placed in a bath of active particles, numerical simulations show the emergence of clusters of passive particles~\cite{mccandlish_spontaneous_2012,stenhammar_activity-induced_2015,Dolai2018,omar_swimming_2019, rodriguez_phase_2020,omar_phase_2021}. Beyond the interest in this fundamental problem, such clustering  may be highly promising for practical applications, such as the realization of assemblies like crystals, gels and micelles from colloidal passive elements~\cite{Kraft2012,SchwarzLinek2012,stenhammar_activity-induced_2015,massana-cid_active_2018,omar_swimming_2019}.

Swimming bacteria are known to transfer part of their activity to passive particles. This has been observed for micron size objects such as beads~\cite{wu_particle_2000,mino_enhanced_2011,valeriani2011colloids,patteson2016particle}, ellipsoids~\cite{peng_diffusion_2016} or more complex shapes such as gears~\cite{leonardo_bacterial_2010,sokolov2010swimming} that exhibit rotation when placed in a bacterial bath. Such an active bath can also mediate short-range effective attractions similar to depletion forces between suspended particles, causing the particles to stick together and form pairs~\cite{angelani_effective_2011} and small clusters~\cite{gokhale2022dynamic}. However, these experiments show limited growth and absence of phase separation, questioning the possibility to form large-scale clusters in a bacterial system.

In this paper, we report continuously growing clusters forming from up to $10^5$ beads suspended in a bacterial bath with apparently no limitation in size. We show that, after a transient governed by the bacteria-induced bead diffusivity, the cluster growth follows a power law $\sim t^{1/3}$,  compatible with the Lifshitz-Slyozov-Wagner theory~\cite{lifshitz1961kinetics,wagner1961theorie,voorhees1985} describing the Ostwald coarsening mechanism for the growth of grains in a supersaturated solution.

\begin{figure}[b]
\includegraphics[width=8.6cm]{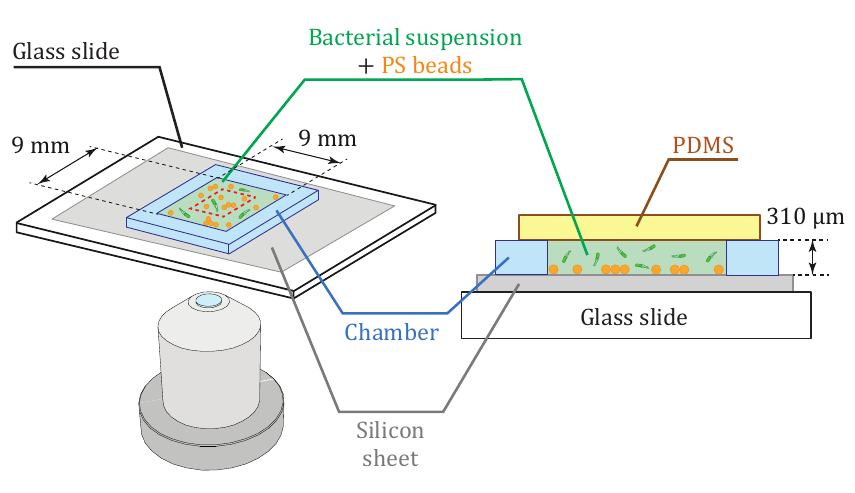} 
\caption{Experimental setup. A drop of a suspension containing polystyrene beads and \emph{B. contaminans} bacteria is placed in a \SI{25}{\uL} frame-sealed chamber (height \SI{310}{\um}, length \SI{9}{\mm}) on a glass slide covered by a silicon sheet to minimize adhesion. The chamber is closed by a PDMS (polydimethylsiloxane) cover to ensure a good oxygenation of the suspension, and placed on the stage of an inverted microscope.}
\label{fig:setup}
\end{figure}

\begin{figure*}[t]
\includegraphics[width=17.5cm, angle=0]{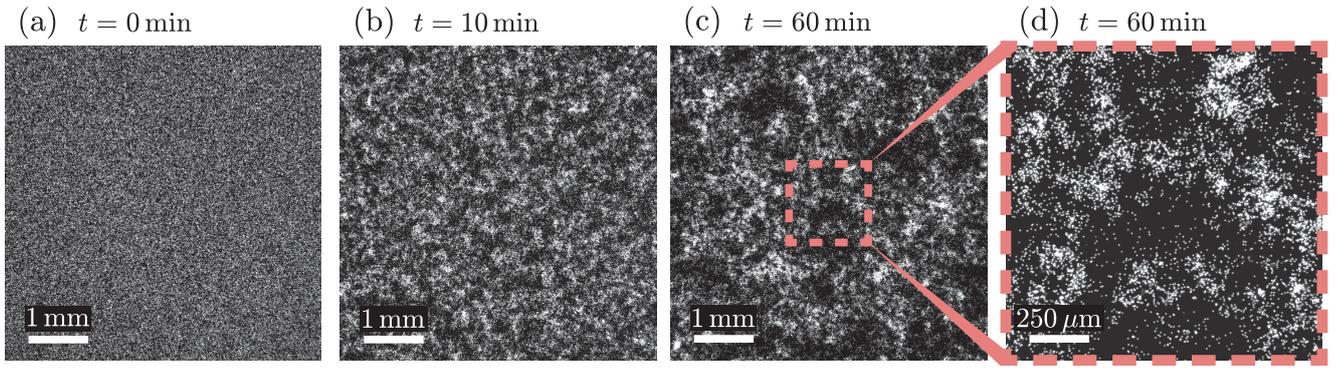} 
\caption{Image sequence showing the clustering of beads of diameter $D_B=\SI{5}{\um}$ and surface fraction $\Phi_B=0.12$  induced by the activity of the bacterial bath at bacterial concentration $\mathrm{OD}=5$ (about $10^5$ beads present on the images). Image (d) is a zoom of image (c), showing freely diffusing beads in the depleted regions between clusters.}
\label{fig:cartoon}
\end{figure*}

\section{Experimental setup}

Our system, sketched in Fig.~\ref{fig:setup}, consists in fluorescent polystyrene beads (PS-FluoGreen) of diameters $D_B$ ranging between 2 and \SI{40}{\um} added to a suspension of {\it{Burkholderia contaminans}}, a motile bacterium ubiquitous in the environment (see Appendix~\ref{sec:SM_bact_prep} for details). The bacterial cell size is approximately \SI{3}{\um} long, \SI{1}{\um} diameter, and the average swimming velocity is $V_s\simeq \SI{20}{\um\per\s}$, with a typical run duration $\tau_\mathrm{run} \simeq  \SI{0.8}{\s}$ and run length $V_s \tau_\mathrm{run} \simeq \SI{16}{\um}$~\cite{Bouvard2022}. A drop of the bead-bacteria mixture is injected in a chamber placed on the stage of an inverted microscope.

The chambers were made by bonding a \SI{25}{\uL} Frame-Sealed Chamber (Bio-Rad), height \SI{310}{\um} and side length \SI{9}{\mm}, on a glass slide covered by a silicon sheet (Gel-Film). This substrate limits the beads adhesion (see Appendix~\ref{sec:SM_kappa} for details). The chamber is closed by a PDMS (polydimethylsiloxane) cover to ensure a good oxygenation of the suspension, and placed on the stage of an inverted microscope (Leica DMI4000B) equipped with a \SI{2048x2048}{pixels} camera (Hamamatsu Orca Flash 4).

Because the beads are denser than the culture medium, they rapidly settle on the bottom surface of the chamber within minutes. The bead concentration is therefore expressed in terms of surface fraction, $\Phi_B$, ranging from \num{3.7e-4} to \num{0.7}. Two bacterial concentrations are used, corresponding to optical densities $\text{OD}=1$ and 5, with $\SI{1}{OD} \sim \SI{1.8e6}{\bact\per\uL}$. These high concentrations ensure an intense swimming activity (and hence a large bead diffusivity), but they remain below the onset of collective motion~\cite{martinez_combined_2020}. 

The large-scale dynamics of the clustering is investigated from image sequences with a large field of view \SI{5.3x5.3}{\mm} and a resolution of \SI{2.6}{\um} per pixel, and images are acquired every $\SI{30}{\s}$ over \SI{1}{h}. We checked that the bacterial  concentration and the ensuing bead activity remain constant during that time (see Appendixes~\ref{sec:SM_control} and \ref{sec:SM_mu} for details). To track individual beads, images are acquired at a higher frame rate, between 1 and \SI{20}{fps}, with a smaller field of view, between \SI{0.21x0.21}{\mm} and \SI{5.3x5.3}{\mm}, chosen according to the bead diameter.

\section{Dynamic clustering}

The sequence of images in Fig.~\ref{fig:cartoon}, obtained for a bead diameter $D_B=\SI{5}{\um}$ and a surface fraction $\Phi_B=0.12$, illustrates the dynamics of the cluster formation (only the fluorescent beads are visible). Starting from uniformly distributed beads at short time (a), denser domains rapidly form at small scale (b), and then gradually evolve toward larger clusters separated by depleted regions (c). A 1-hour time-lapse of this experiment (see \href{http://www.fast.u-psud.fr/~auradou/Bacteries-Beads/Movie1.avi}{Suppl. Movie 1}) illustrates the clustering dynamics, showing dense concentrations of beads highly fluctuating in size and continuously exchanging beads. Additional movies with higher magnification and acquisition rates (see \href{http://www.fast.u-psud.fr/~auradou/Bacteries-Beads/Movie2.avi}{Suppl. Movie 2}, \href{http://www.fast.u-psud.fr/~auradou/Bacteries-Beads/Movie3.avi}{Suppl. Movie 3}) emphasize this continuous exchange of beads between clusters, with freely diffusing beads in the depleted areas between the clusters.

We observe that the clustering dynamics remains essentially two-dimensional during most of the clustering process, except in the densest regions at large time, where beads occasionally overlap in two or more layers. Experiments performed without bacteria or with non-motile (bleach-inactivated) bacteria do not show any clustering, confirming that the clustering dynamics is driven by the swimming of bacteria. Additional experiments performed with wild type \emph{E. coli} show a similar clustering dynamics but on a slower time scale (see Appendix~\ref{sec:SM_ecoli} for details). This results in smaller clusters at the end of the experiments, which led us to favor experiments with \emph{B. contaminans}.

The clustering dynamics, similar to that observed with self-propelled particles~\cite{buttinoni_dynamical_2013,stenhammar_phase_2014,stenhammar_activity-induced_2015,cates_motility-induced_2015,wysocki_cooperative_2014,Bialke2015,Ginot2018,omar_phase_2021,van_der_linden_interrupted_2019,Paoluzzi2022}, is reminiscent of the Ostwald ripening mechanism~\cite{lifshitz1961kinetics,wagner1961theorie,voorhees1985}. In the Ostwald ripening model, individual constituents are subject to a short-range attraction, and an excess of concentration around the clusters, proportional to their curvature, generates a diffusive flux from small to large clusters. In other words, the constituents ``evaporate'' from small clusters and ``condense'' on larger clusters, leading to a growth as $t^{1/3}$ at large time~\cite{lifshitz1961kinetics,wagner1961theorie,stenhammar_phase_2014,cates_motility-induced_2015,zhang2021active}. To check to what extent this picture holds in the present study, we quantify in the following the time evolution of the cluster size, and determine the bacteria-induced attractive force between beads from bead tracking measurements.

To characterize the cluster size, we have developed a robust image-based method, applicable even when the beads are too small to be resolved individually (see Appendix~\ref{sec:SM_cluster_size} for details). We define an image heterogeneity index as
\begin{equation}
\label{eq:sigma_n}
\sigma_n(t,L_{F}) = \dfrac{\sigma(I({\bf x}, t)*G_{L_F}({\bf x}))}{\sigma(I({\bf x}, t))}, 
\end{equation}
i.e., as the standard deviation of the intensity levels $I({\bf x}, t)$ of the image at time $t$ convoluted with a Gaussian filter $G_{L_F}(\bf x)$ of width $L_F$, normalized such that $\sigma_n(0,t)=1$.  With this definition, one has $\sigma_n(L_F,t) \rightarrow 0$ when the filter width $L_F$ exceeds the largest scale present in the image. We finally define the cluster size $L_c(t)$ as the filter width such that $\sigma_n(L_c(t))=1/2$.

\begin{figure}[tb]
\centering
  \includegraphics[width=8cm]{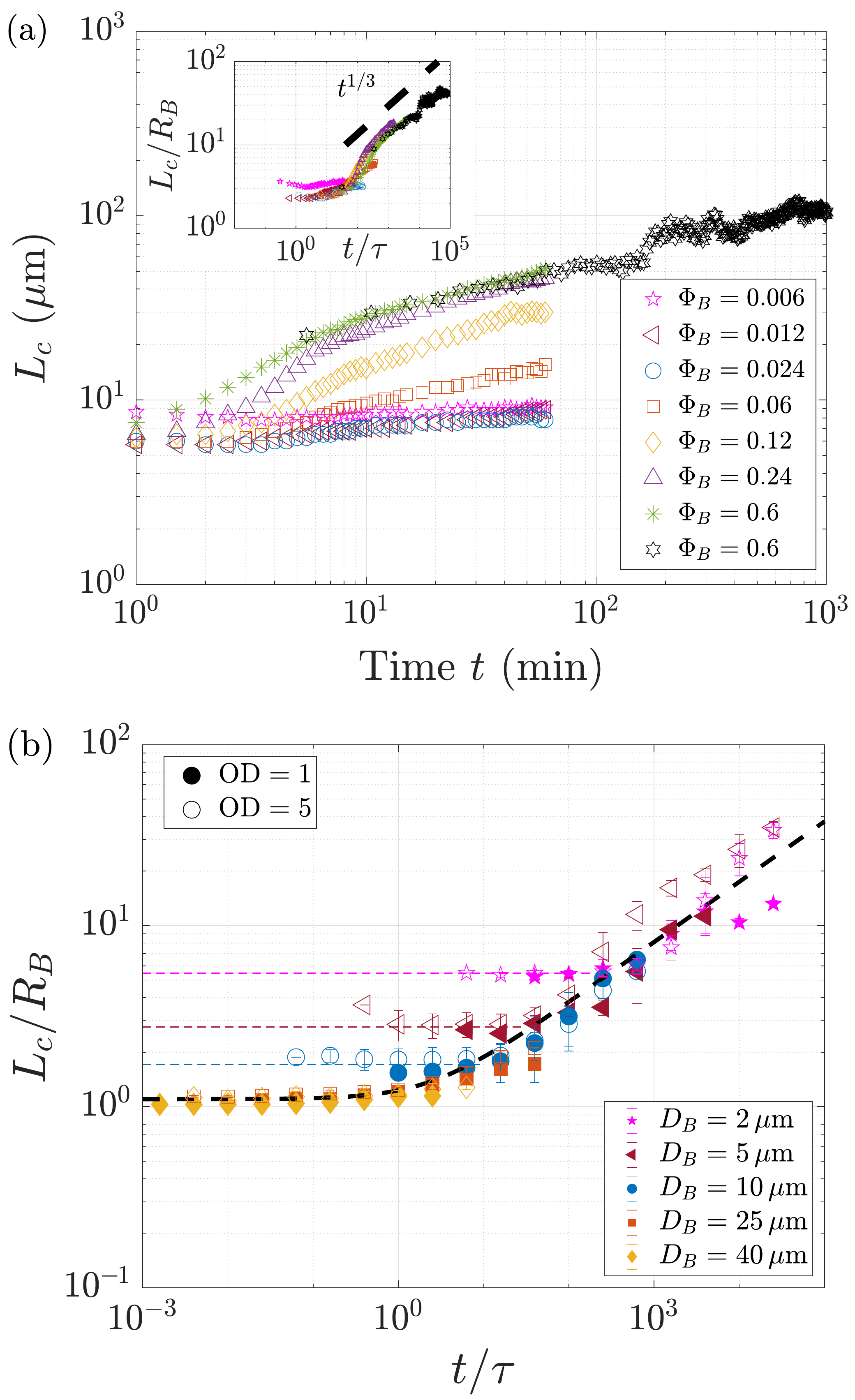}
  \caption{(a) Cluster size $L_c$ as a function of time at large bacterial concentration ($\mathrm{OD}=5$) for beads of diameter $D_B=\SI{5}{\um}$ and various surface fractions $\Phi_B$. Inset: Normalized cluster size $L_c/R_B$ as a function of $t/\tau$ for $D_B=\SI{5}{\um}$. (b)  $L_c/R_B$ as a function of $t/\tau$ for various bead diameters $D_B$ and bacterial concentrations (filled symbols: $\mathrm{OD}=1$; empty symbols: $\mathrm{OD}=5$). The vertical bars reflect the dispersion for each bead diameter. The black dashed line shows a best fit using Eq.~(\ref{eq:model2}) with $\alpha=\num{0.4\pm0.2}$ and $\beta=\num{1.1\pm0.1}$. The colored dashed lines emphasize the larger apparent cluster size for small bead diameters due to the diffraction limit.}
  \label{fig:Lc_t}
\end{figure}

Figure~\ref{fig:Lc_t}(a) shows the temporal evolution of the cluster size $L_c$ for beads of diameter \SI{5}{\um} at various surface fractions $\Phi_B$. For the lowest $\Phi_B$, $L_c$ remains close to the bead diameter, and shows only a slight increase after about ten minutes, corresponding to small clusters made of $2-4$ beads only. As the surface fraction is increased, the transition to the growth regime starts earlier, and the cluster size increases by a factor of almost 10 after 1 hour, indicating clusters made of typically 100 beads, with no visible saturation.

\begin{figure}[b]
\centering
  \includegraphics[width=7.5cm]{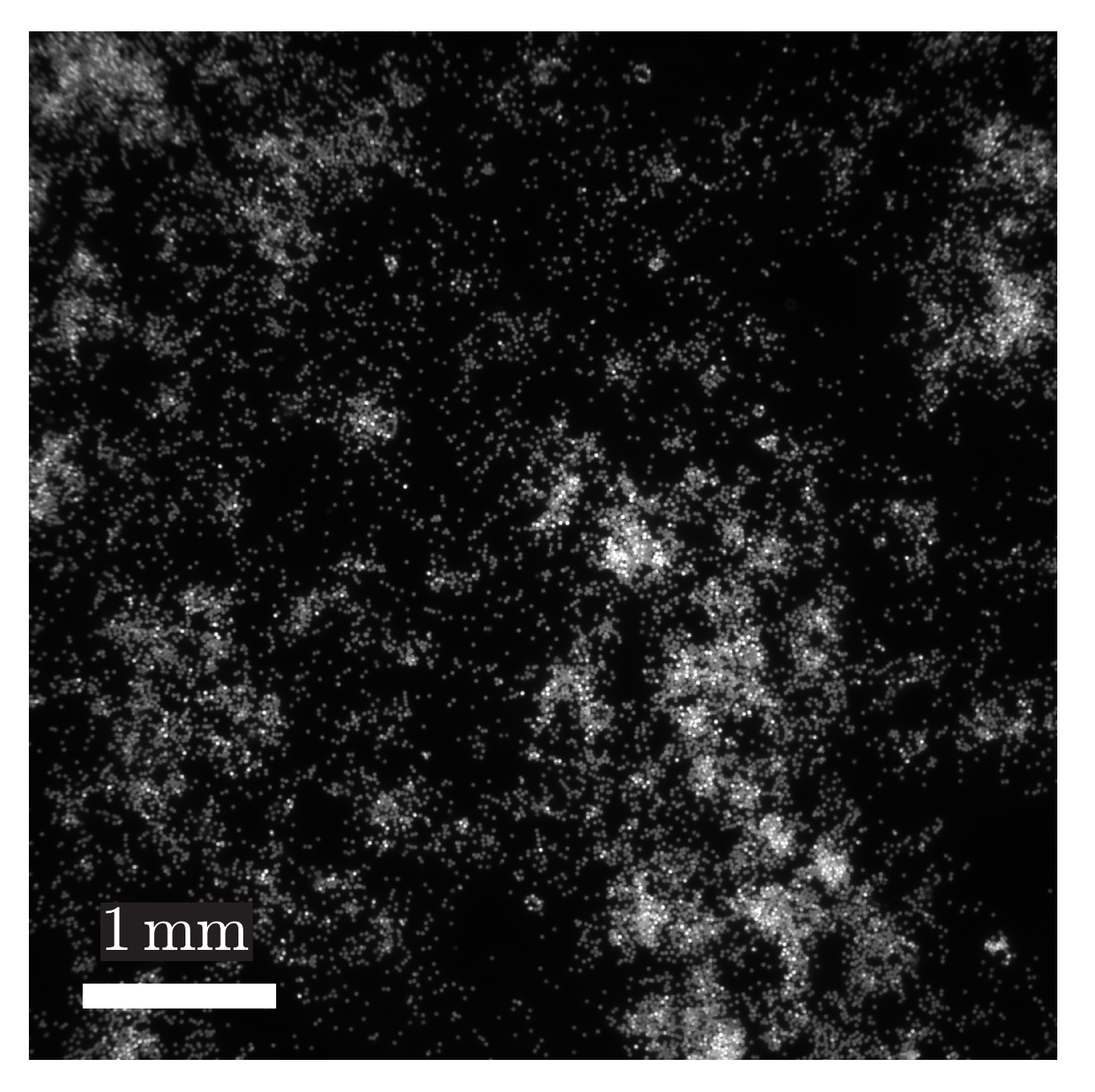}
  \caption{Giant bead clustering after 16 hours, obtained for beads of diameter $D_B=\SI{5}{\um}$ and initial bacterial concentration of $\mathrm{OD}=5$ and surface fraction $\Phi_B=0.6$.}
  \label{fig:snap16h}
\end{figure}

To examine the robustness of this growth at large time, we performed an additional experiment lasting \SI{16}{h} in the particular case of $D_B=\SI{5}{\um}$, $\mathrm{OD}=5$, and $\Phi_B=0.6$ [black symbols in Fig.~\ref{fig:Lc_t}(a)]. Although the bacterial density is not constant over this long duration, this test experiment confirms the continuous increase of $L_c$, up to a factor of 30 at the end of the experiment, corresponding to clusters made of $\sim 10^3$ beads. The final image after 16~hours (Fig.~\ref{fig:snap16h}) shows large clusters, characterized by a cluster size $L_c = \SI{150}{\um}$. These clusters are themselves organized in structures of an even larger size, up to $\SIrange{2}{3}{\mm}$, making them visible to the naked eye.

\begin{figure}[tb]
\centering
  \includegraphics[width=8.6cm]{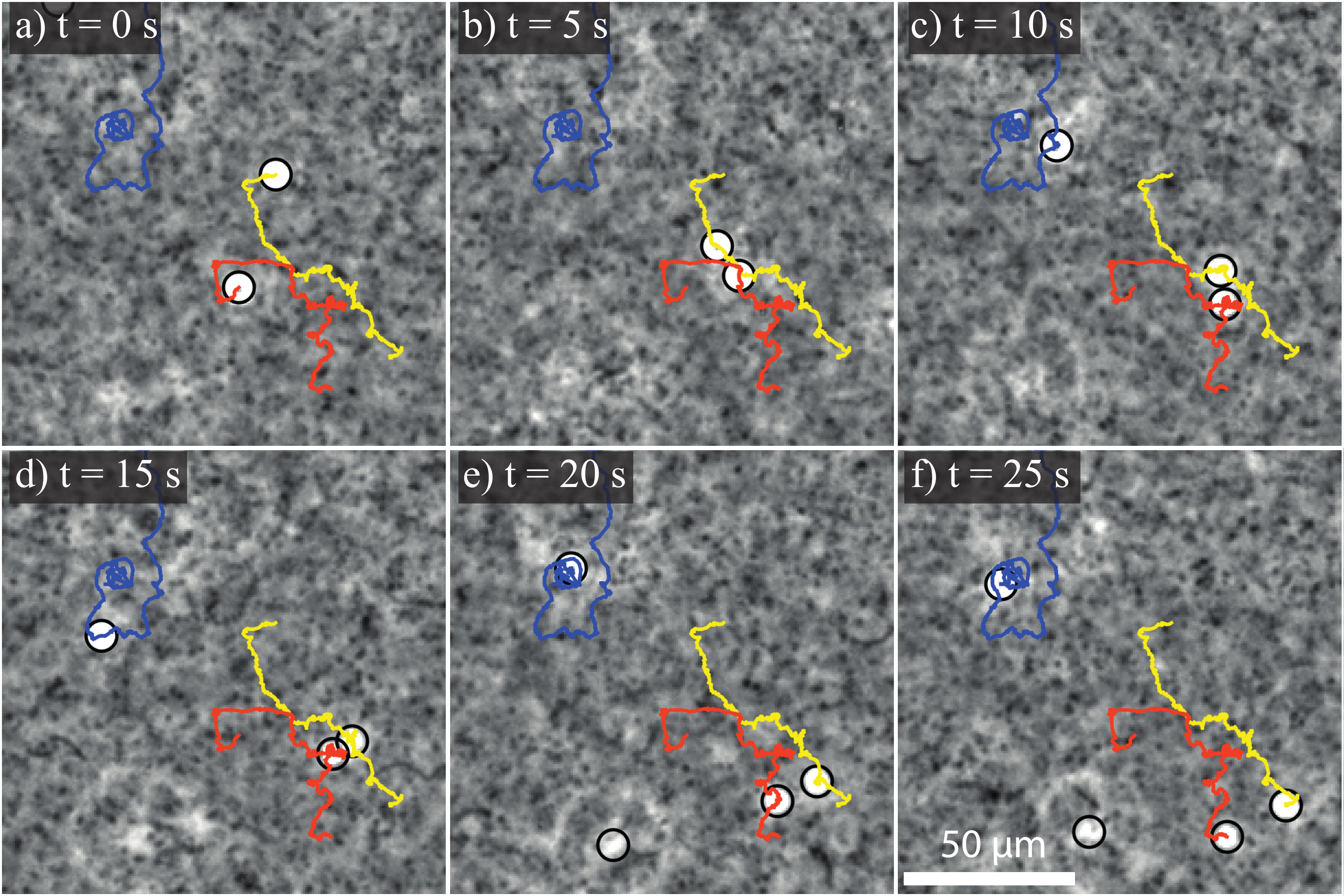}
  \caption{Sequence of phase-contrast images taken from the \href{www.fast.u-psud.fr/~auradou/Bacteries-Beads/Movie4.avi}{Suppl. Movie 4} showing a freely diffusing bead (blue track) and two interacting beads (yellow and red tracks). Bead diameter $D_B=\SI{5}{\um}$, bacterial concentration $\mathrm{OD}=5$.}
  \label{fig:SM_contact_video}
\end{figure}

\begin{figure}[b]
\centering
  \includegraphics[width=8cm]{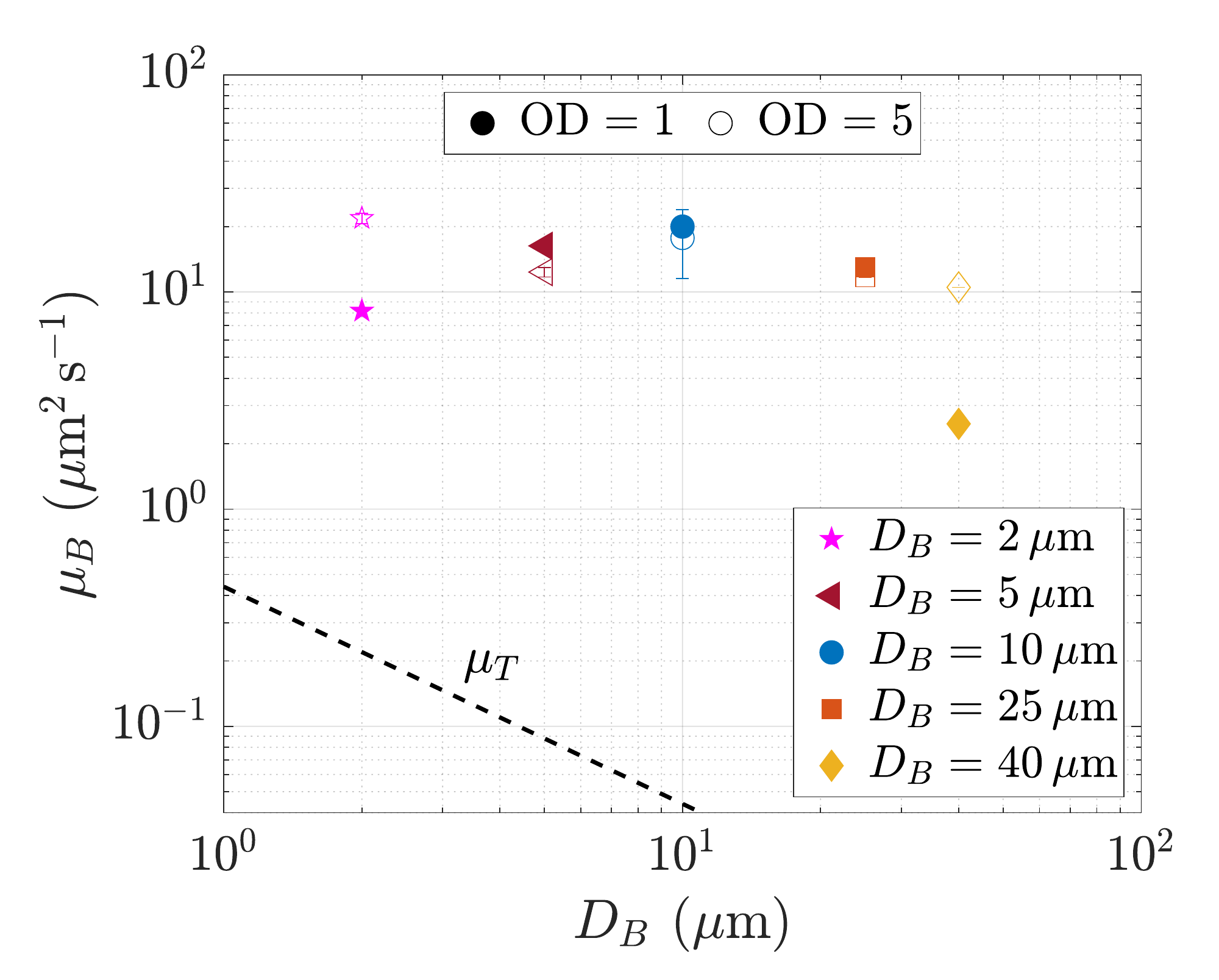}
  \caption{Effective diffusion coefficient $\mu_B$ of the beads as a function of their diameter $D_B$ for bacterial concentrations $\mathrm{OD}=1$ (filled symbols) and $\mathrm{OD}=5$ (empty symbols). The dashed line shows the Stokes-Einstein prediction $\mu_T=k_B T/(6\pi\eta R_B)$ for thermal diffusion.}
  \label{fig:SM_Diffusion}
\end{figure}

The earlier onset of clustering at larger surface fraction observed in Fig.~\ref{fig:Lc_t}(a) suggests that the clustering time is governed by the time of first encounter of freely diffusing beads. Accordingly, this crossover time must be governed by the bead concentration and the bacteria-induced bead diffusion coefficient $\mu_B$. To estimate $\mu_B$, we track isolated beads in experiments performed with a lower surface fraction, as illustrated in Fig.~\ref{fig:SM_contact_video}. We compute the mean square displacement
\begin{equation}
    \langle |\delta \mathbf{r}|^2 \rangle = \langle |{\mathbf{r}}_i(t)-\mathbf{r}_i(0)|^2\rangle,
\end{equation}
where $\mathbf{r}_i(t)$ is the position of the bead $i$ and the brackets $\langle\cdot\rangle$ is the average over time and beads (see Appendix~\ref{sec:SM_mu} for details). In the diffusive regime (large times), we observe a linear dependence, $\langle |\delta {\bf r}|^2 \rangle \simeq 4 \mu_B t$, from which we compute $\mu_B$. Figure~\ref{fig:SM_Diffusion} summarizes the values of $\mu_B$ for various bead diameters and the two bacterial concentrations. We obtain $\mu_B$ between 2 and \SI{25}{\square\um\per\s}, with a weak non-monotonic dependence on the bead diameter. This bacteria-induced diffusivity is $10^2$ to $10^3$ larger than the thermal diffusivity expected for beads of the same diameter shown in dotted line, $\mu_T=k_B T/(6\pi\eta R_B)$, with $k_B$ the Boltzmann constant, $T$ the temperature and $\eta$ the fluid viscosity~\cite{einstein1905molekularkinetischen}. This strong diffusivity, in agreement with previous studies~\cite{wu_particle_2000,mino_enhanced_2011,patteson2016particle}, highlights the strong bead-bacteria coupling for bead diameters of the order of the swimming persistence length $V_s \tau_\mathrm{run}$.

From this bacteria-induced bead diffusivity, we finally estimate the characteristic time of first encounter, $\tau = \ell^2/\mu_B$, for two beads separated by a distance $\ell$. At short time, the beads are randomly distributed on the bottom surface of the cell, so that $\ell \simeq R_B/\sqrt{\Phi_B}$ with $R_B$ the bead radius, yielding
\begin{equation}
\label{eq:tau}
\tau \simeq \frac{R_B^2}{\Phi_B \mu_B}.
\end{equation}
Normalizing the time $t$ with $\tau$ and the cluster size $L_c$ with $R_B$ yields a good collapse of our data at $D_B=\SI{5}{\um}$ and $\mathrm{OD}=5$ for various surface fractions [see inset of Fig.~\ref{fig:Lc_t}(a)], confirming that the onset of clustering is governed by $\tau$.

At short time, in the freely diffusing bead regime ($t \ll \tau$), the data plateau to a constant value. This plateau takes the expected value $L_c/R_B \simeq 1$ for the largest beads, but is significantly larger for $D_B \leq \SI{10}{\um}$. This originates from the diffraction limit of optical microscopy~\cite{Wolf2007}: for beads smaller than the microscope resolution, the apparent cluster size is given by the Airy diffraction diameter rather than the true bead diameter, yielding an overestimation of the apparent cluster size. 

At large time, in the clustering regime ($t \gg \tau$), the data approximately follow a power law with an exponent consistent with $1/3$ (dotted line),  in agreement with the Lifshitz-Slyozov-Wagner prediction~\cite{lifshitz1961kinetics,wagner1961theorie,bray_theory_2002}, with no apparent saturation. These results are confirmed by the experiments performed with various bead sizes ($D_B=\SIrange{2}{40}{\um}$) and two bacterial concentrations [see Fig.~\ref{fig:Lc_t}(b)]. In this figure, each data set is averaged over 4 to 8 experiments, with vertical bars reflecting the dispersion between experiments. At large time, $t / \tau > 100$, all data follow a power-law growth $L_c \sim t^a$ with $a \simeq \num{0.35\pm0.05}$.

\section{Bacteria-induced attractive force}

\begin{figure}[tb]
\includegraphics[width=8cm]{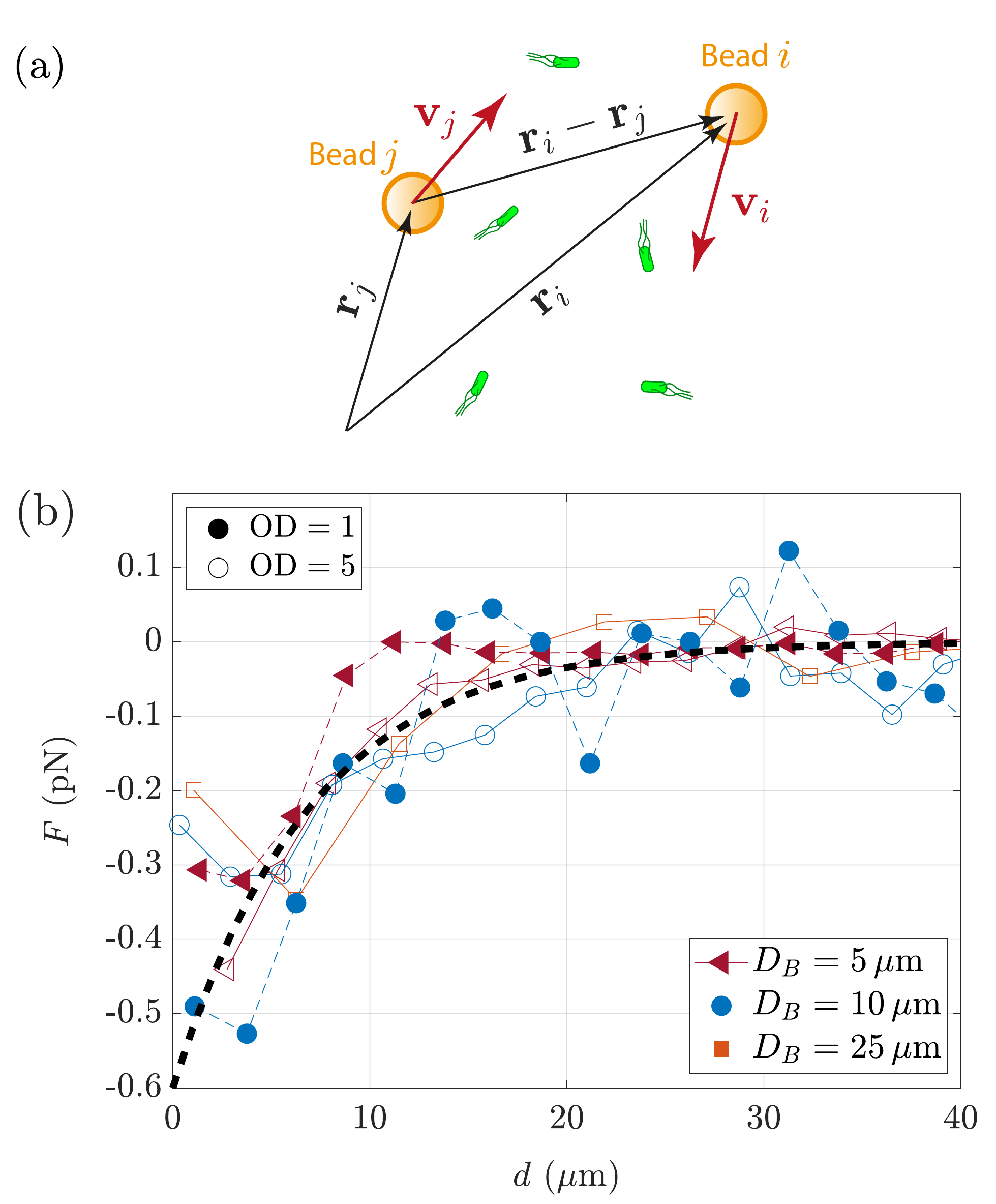} 
\caption{(a) Sketch of the velocity difference $\delta V_{ij}$ for a pair of beads $(i,j)$. (b) Average force $F$ between two beads as a function of their separation $d$, for beads of diameter $D_B=5, 10, \SI{25}{\um}$ at $\mathrm{OD}=1$ and 5. Black dashed line: exponential fit $F(d)=-F_0 \, e^{-d/d_0}$ with $F_0=\SI{0.6\pm0.2}{pN}$ and $d_0=\SI{7\pm1}{\um}$.}
\label{fig:force}
\end{figure}

We now address the key question of the  effective bacteria-induced attractive force at the origin of the large-scale clustering of the beads. The beads being essentially freely diffusing at short time suggests that this attractive force acts only when the beads are very close. To characterize this short-range attraction, we analyze the relative motion of bead pairs at low surface fraction $\Phi_B$ and high acquisition rate (see Fig.~\ref{fig:SM_contact_video} and \href{www.fast.u-psud.fr/~auradou/Bacteries-Beads/Movie4.avi}{Suppl. Movie 4} for details). For each pair of beads $(i,j)$, we compute their velocity difference projected along the pair separation [see Fig.~\ref{fig:force}(a)],
\begin{equation}
\label{eq:velocity}
    \delta V_{ij} = (\mathbf{v}_i - \mathbf{v}_j) \cdot \dfrac{\mathbf{r}_i-\mathbf{r}_j}{\vert\mathbf{r}_i-\mathbf{r}_j\vert}.
\end{equation}
Two contributions are expected in $ \delta V_{ij}$: a large stochastic contribution reflecting the bacteria-induced diffusivity, which should average to zero, and a small deterministic negative contribution reflecting the attraction. We extract this deterministic contribution by computing  the relative velocity $\overline{\delta V}(d)$ averaged over the pairs $(i,j)$ separated by a given distance $d=\vert\mathbf{r}_i-\mathbf{r}_j\vert-2R_B$. To deduce the bacteria-induced effective force, we assume that this force is balanced by the viscous drag, which writes in the form
\begin{equation}
\label{eq:kappa}
F(d) = - \kappa \eta R_b \, \overline{\delta V}(d),
\end{equation}
with $\eta$ the liquid viscosity and $\kappa$ a friction factor that characterizes the motion of the bead over the bottom surface of the chamber ($\kappa = 6\pi$ for an isolated sphere far from the walls). We determine empirically $\kappa$ by measuring the velocity of beads drifting in a chamber tilted by a small angle (see Appendix~\ref{sec:SM_kappa} for details).

The resulting force, plotted in Fig.~\ref{fig:force}(b), shows a clear attraction at short range, up to $d\simeq\SI{15}{\um}$, and fluctuations around zero at larger $d$, as expected for pure diffusion. An increase as $d \rightarrow 0$ is also observed, corresponding to the steric interaction between beads. To extract the characteristic range of the attraction, we fit the attractive part of $F$ with $F(d) \simeq -F_0 e^{-d/d_0}$. This fit gives a length scale $d_0 \simeq 7 \pm 1\,\si{\um}$ of the order of the bacteria size (including flagella), with weak dependence on the bead diameter. The characteristic force is $F_0 \simeq  0.6 \pm 0.2$\,pN, a value close to the hydrodynamic force induced by the swimming of bacteria~\cite{Drescher2011}. This force magnitude and range are much larger than those expected for the van der Waals force for polystyrene beads~\cite{ohshima2016van,haliyooverview}, confirming that the swimming activity of bacteria is the dominant process in the bead clustering.

We finally discuss to what extent the Ostwald ripening mechanism describes the observed cluster growth. In the classical Ostwald ripening, the "evaporation-condensation" quasi-equilibrium state is governed by the ratio between the short-range attractive potential $E_0$ and the kinetic energy $k_B T$. In our system, an effective potential can be estimated from the bacteria-induced attractive force as $E_0 \simeq F_0 d_0$, and an effective kinetic temperature can be defined from the bead diffusivity  using a Stokes-Einstein law modified by the wall-bead friction factor (\ref{eq:kappa}), $k_B T_\mathrm{eff} \simeq \kappa \eta R_B \mu_B$. We obtain a ratio $\sigma = E_0 / k_B T_\mathrm{eff}$ ranging from 0.15 for the \SI{25}{\um} beads, which slowly form compact clusters, and 1.6 for the \SI{5}{\um} beads, which rapidly form loosely packed and highly fluctuating clusters. According to the Ostwald mechanism, the deficit of bounding energy for beads in small clusters generates a small excess of concentration of free beads proportional to the cluster curvature, $\Delta \Phi_B \simeq \sigma R_B/L_c$, leading to a preferential diffusion toward larger clusters.  This results in a self-similar growth given by the Lifshitz-Slyozov-Wagner relation~\cite{lifshitz1961kinetics,voorhees1985,wagner1961theorie},
\begin{equation}
    \label{eq:model2}
    \dfrac{L_c}{R_B} = \beta \left(1 + \alpha \dfrac{t}{\tau}\right)^{1/3},
\end{equation}
with $\alpha$ and $\beta$ of the order of unity. Our data is well described by this law for large beads [see Fig.~\ref{fig:Lc_t}(b)], with $\beta \simeq \num{1.1\pm0.1}$ and $\alpha \simeq \num{0.4\pm0.2}$.  The data for smaller beads show a higher plateau at short time due to the diffraction limit, but collapse on the predicted power-law growth at large time. We therefore conclude that the Ostwald ripening mechanism provides a good description of the bacteria-induced bead clustering in our system.

\section{Conclusion}

In this paper, we show that the swimming activity of bacteria produces a clustering of passive beads over a wide range of sizes and concentrations, following an Ostwald ripening-like dynamics with no apparent saturation. Our experiments demonstrate that very large clusters can be formed, raising the possibility of using bacterial motility to generate large structures of passive colloidal particles.

Our ever-growing clusters differ from the small quasi-stationary clusters recently reported by Gokhale {\it et al.}~\cite{gokhale2022dynamic}. This difference may originate from the much larger bead diffusivity measured in our system, resulting in a shorter time of first encounter and hence a faster dynamics. Several parameters may explain this larger diffusivity, among which the use of a highly motile bacterial strain, a weak vertical confinement, and a strongly reduced wall-bead friction achieved by adding a silicone substrate on the bottom of the chamber. 

Our experiments allowed us to provide an estimate for the bacteria-induced attractive force between beads. This force may combine a steric effect originating from the depletion of bacteria between close beads, similar to particles suspended in a solution of macromolecules~\cite{Harder2014,Ray2014,Ni_Cohen,omar_phase_2021,asakura_interaction_1954,asakura_interaction_1958}, and a hydrodynamic effect, originating from the viscous stress induced by the swimming of bacteria. Close to a solid boundary, this stress tends to maintain bacteria on the surface~\cite{berke_hydrodynamic_2008,Sipos2015}, yielding an attractive force between beads. This picture is confirmed by our force estimate, of magnitude matching the hydrodynamic thrust of a swimming of bacteria and range of the order of typical bacteria size~\cite{Drescher2011}.

\section*{Acknowledgements}

We thank A. Gargasson for experimental help, and G. Dietze and M. Jarrahi for fruitful discussions.
This work is supported by the French National Research Agency (ANR) through the "Laboratoire d'Excellence Physics Atom Light Mater" (LabEx PALM) as part of the "Investissements d'Avenir" program (ANR-10-LABX-0039).


\appendix

\section{Bacterial strains and culture}
\label{sec:SM_bact_prep}

The main bacterial strain used in this study is {\it{Burkholderia contaminans}}, an environmental strain characterized from the sequencing of {\it 16S} and {\it recA} gene fragments~\cite{bouvard2022thesis,Bouvard2022}. Bacteria were grown overnight in YEB (Yeast Extract Beef) medium at \SI{28}{\celsius} in an orbital shaker at \SI{200}{rpm}. A certain volume of this suspension is added to YEB containing fluorescent polystyrene beads (PS-FluoGreen, from microParticles GmbH). The volume is chosen to obtain a suspension containing $1.8$ or \num{9e6} bacteria per microliter, corresponding to an optical density (OD) of 1 and 5, respectively. The solution is then homogenized with an orbital vortex mixer and placed in the chamber.

Additional experiments were conducted with another bacterial species, \emph{Escherichia coli} RP 437. The bacteria were cultured in CAM-M9G (1:1000 ratio) medium in an incubator shaker at \SI{240}{rpm} and \SI{30}{\celsius}. The culture was stopped at an OD close to 0.1 and washed twice by centrifugation (\SI{2300}{g} for \SI{10}{min}). The bacteria were then re-suspended into a minimal buffer (0.1\,M EDTA, 0.001\,M Methionine, 1\,M Sodium Lactate and 0.1\,M Phosphate buffer dissolved in milliQ water at pH = 7.0). This medium provides the salts needed for the bacterial swimming, but it does not contain any nutrient for bacterial growth. The bacteria were concentrated at an OD of 7 and then diluted to $\mathrm{OD}=5$, before adding \SI{2}{\um} fluorescent beads.

\section{Measurement of the characteristic cluster size}
\label{sec:SM_cluster_size}

\begin{figure}[b]
\begin{center}
\includegraphics[width=8cm]{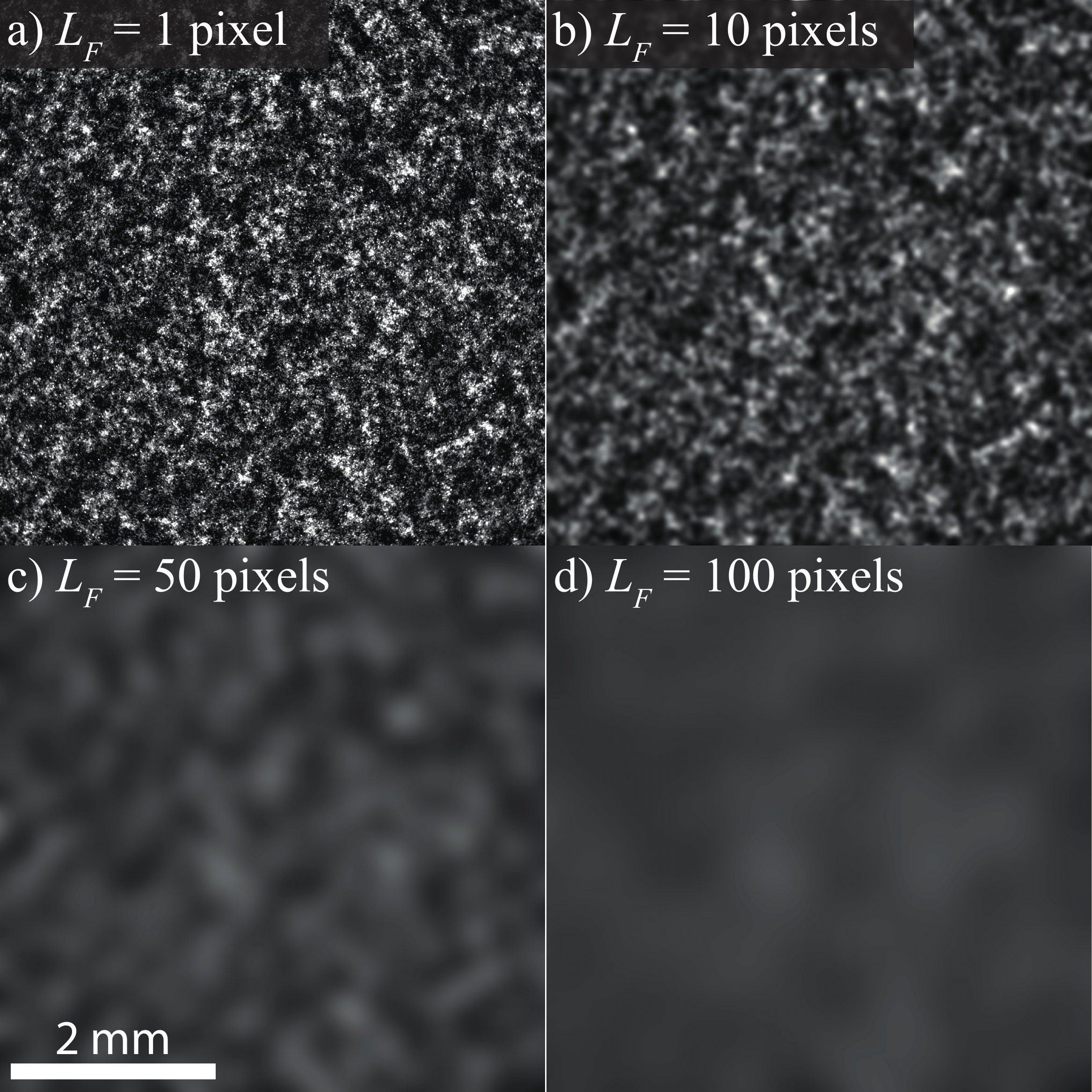}
\caption{(a) Image (size $2048 \times 2048$ pixels) of fluorescent polystyrene beads of diameter $D_B=\SI{5}{\um}$ with surface fraction $\Phi_B=0.24$, in a suspension of \emph{B. contaminans}. (b,c,d) Same image convoluted with a Gaussian filter $G_{L_F}(\bf x)$ of width $L_{F}=10$, 50 and 100\,pixels, used to determine the characteristic cluster size.}
\label{fig:SM_gaussian_blur}
\end{center}
\end{figure}

To compute the characteristic size of the clusters $L_c$, the images are blurred with a low pass Gaussian filter of varying width (see Fig.~\ref{fig:SM_gaussian_blur}). When the filter width exceeds the characteristic size of the clusters, the blurred image becomes nearly uniform, which provides an estimate of the cluster size. This method applies for all bead sizes and magnifications, even for small beads under the microscope resolution.  

\begin{figure}[tb]
\centering
  \includegraphics[width=8cm]{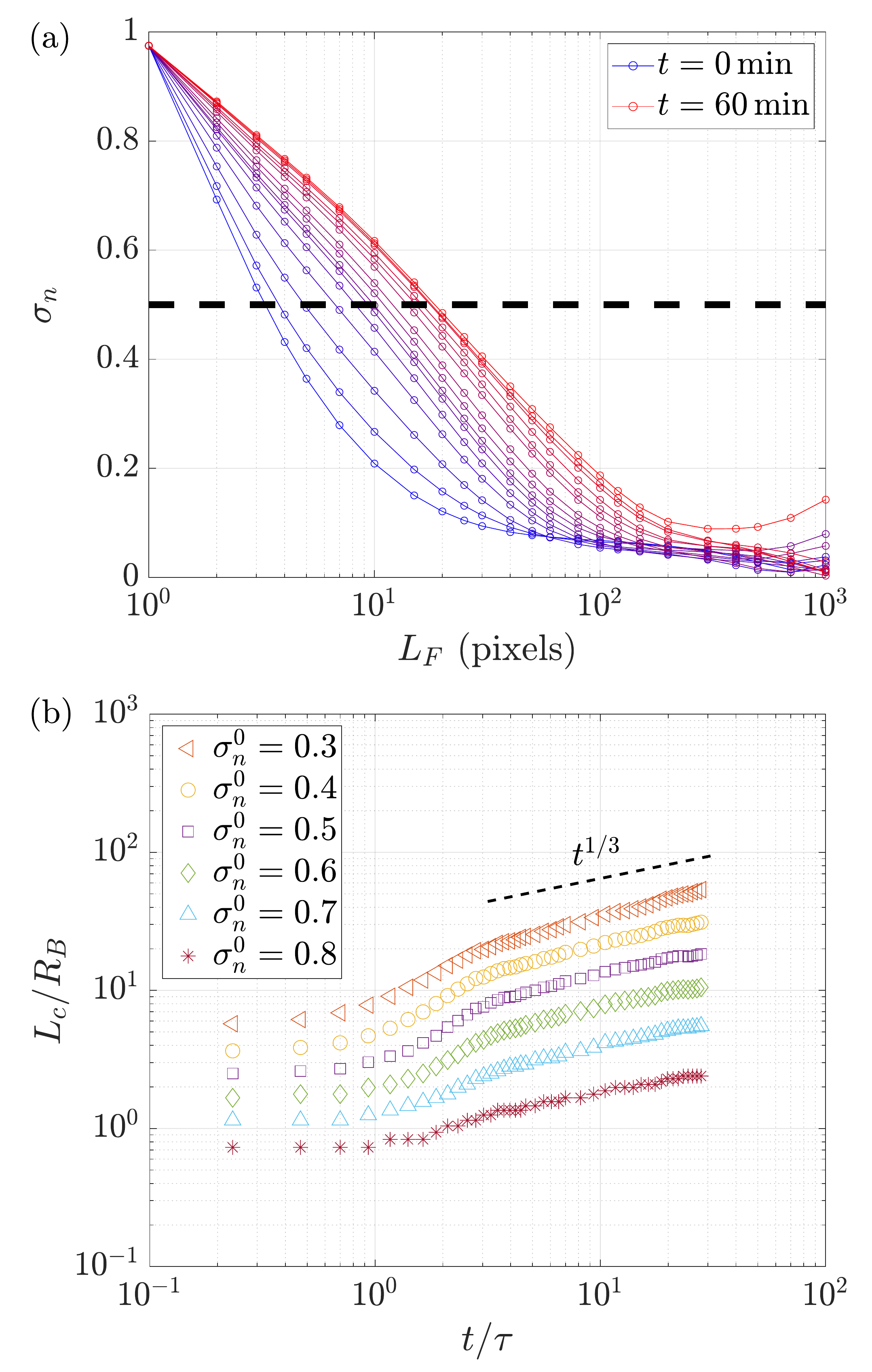}
  \caption{(a) Heterogeneity index $\sigma_n$ as a function of the filter size $L_F$ in pixels ($\SI{1}{pixel}=\SI{2.6}{\um}$). Data obtained for beads of diameter $D_B=\SI{5}{\um}$ at a surface fraction $\Phi_B=0.24$ in a bacterial suspension at $\mathrm{OD}=5$. $\sigma_n(L_F)$ is plotted every $\delta t=\SI{90}{\s}$  from $t=0$ (blue) to $t=\SI{60}{min}$ (red). The black dashed line shows the threshold $\sigma_n^0=0.5$ defining the cluster size $L_c(t)$.
  (b) Evolution of the normalized characteristic cluster size $L_c/R_B$, as a function of the normalized time $t/\tau$ for various values of the threshold $\sigma_n^0$.}
  \label{fig:SM_sigma_n}
\end{figure}

The blurring is achieved by convoluting each image $I({\bf x}, t)$ (of size $2048\times 2048$ pixels) with a Gaussian kernel $G_{L_F}({\bf x})$ of width $L_F$ increasing from 1 to \SI{1000}{pixels}. The standard deviation $\sigma(I({\bf x}, t) * G_{L_F}({\bf x}))$ of the resulting image is then computed. We define the heterogeneity index as the ratio between this standard deviation and the standard deviation of the initial image,
\begin{equation}
\label{eq:sigma_n}
\sigma_n(t,L_{F}) = \dfrac{\sigma(I({\bf x}, t)*G_{L_F}({\bf x}))}{\sigma(I({\bf x}, t))}. 
\end{equation}
This quantity is 1 for the original (non-blurred) image, and tends to 0 for $L_F \gg L_c$ (uniform image), where $L_c$ is the characteristic size of the clusters. Because of the finite size of the image, and because of residual heterogeneity in the illumination, $\sigma_n$ tends to a small value $\sigma_n^*$ of the order of 0.05, which we subtract from $\sigma_n$.

Figure~\ref{fig:SM_sigma_n}(a) shows the heterogeneity index as a function of the filter width $L_F$ at increasing times during the clustering process. The decrease of $\sigma_n$ is shallower at large time, indicating the presence of patterns of increasing size. We define the cluster size $L_c(t)$ as the filter size for which  $\sigma_n(L_c(t))$ reaches a chosen threshold $\sigma_n^0$. The time evolution of $L_c(t)$ obtained for various thresholds $\sigma_n^0$ is illustrated in Fig.~\ref{fig:SM_sigma_n}(b), in the case $\mathrm{OD}=5$, $D_B = \SI{5}{\um}$ and $\Phi_B = 0.24$. All curves show a similar behavior, with a transition from a constant value at short time to a power-law growth $\sim t^{1/3}$ at large time. This indicates that the heterogeneity index (\ref{eq:sigma_n}) provides a robust description of the dynamics of the cluster growth.  The threshold $\sigma_n^0 = 0.5$ is chosen because for large beads (when diffraction effects can be neglected), it yields $L_c \simeq R_B$ at short time.

Other image processing methods were tested, such as the image correlation method, and gave similar results~\cite{bouvard2022thesis}. The present method was found to be more robust to the presence of defects on the images, such as scratches on the glass slide or non-uniform illumination.

\section{Control experiments}
\label{sec:SM_control}

\begin{figure}[b]
\centering
  \includegraphics[width=8cm]{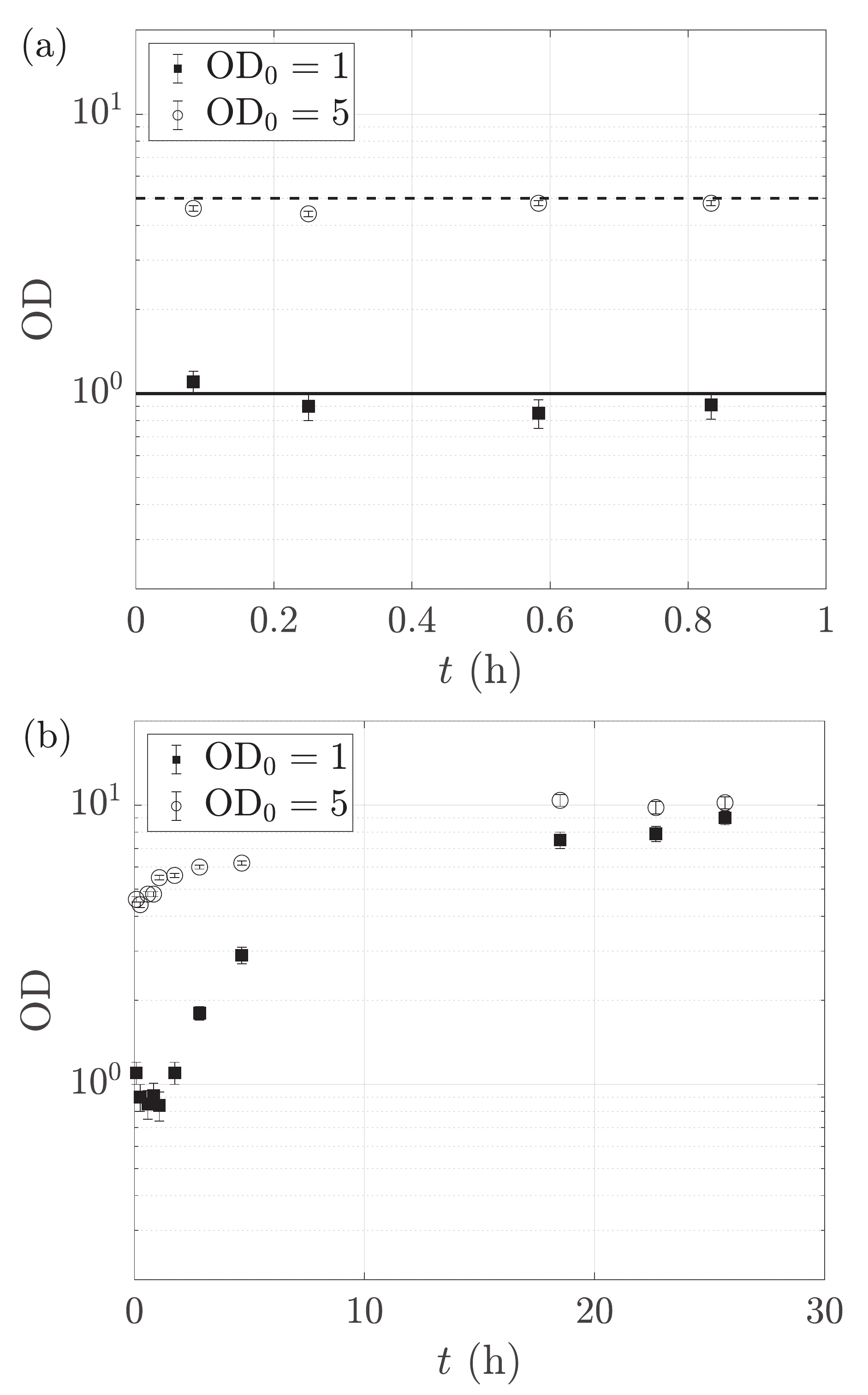}
  \caption{Time evolution of the bacterial concentration of \emph{B. contaminans} of initial optical density 1 (filled squares) and 5 (empty circles). The suspensions were kept under conditions close to those of the experiments. (a) OD values measured during the first hour, corresponding to the duration used in the clustering experiments. (b) OD values measured over twenty-five hours. These measurements show that the OD saturates to 10 beyond ten hours.}
  \label{fig:SM_OD}
\end{figure}

Two experiments were conducted to check that the clustering of beads is induced by the swimming activity of the bacteria. First, a suspension containing only beads and no bacteria did not show any clustering. Secondly, we used a suspension of beads mixed with bleach-inactivated bacteria. Again, no clustering of the beads was observed here.

To check that the bacterial concentration was not changing significantly during the experiments, two suspensions with an initial optical density (OD) of 1 and 5 were placed in conical flasks and kept at room temperature to reproduce the conditions under which the suspensions are maintained during the clustering experiments. Figure~\ref{fig:SM_OD}(a) shows the OD measured on samples taken during one hour.
For both experiments, we can see that the OD remains almost constant over the whole duration of the sampling. Beyond one hour, the bacterial concentration increases and saturates at an OD close to 10 [see Fig.~\ref{fig:SM_OD}(b)]. These measurements indicate that the bacterial concentration for the 16-hour experiment performed at initial $\mathrm{OD}=5$ [shown in Fig.~\ref{fig:Lc_t}(a)] almost doubled over the duration of the experiment.

\section{Clustering experiments with \emph{E. coli}}
\label{sec:SM_ecoli}

\begin{figure}[t]
\centering
  \includegraphics[width=8cm]{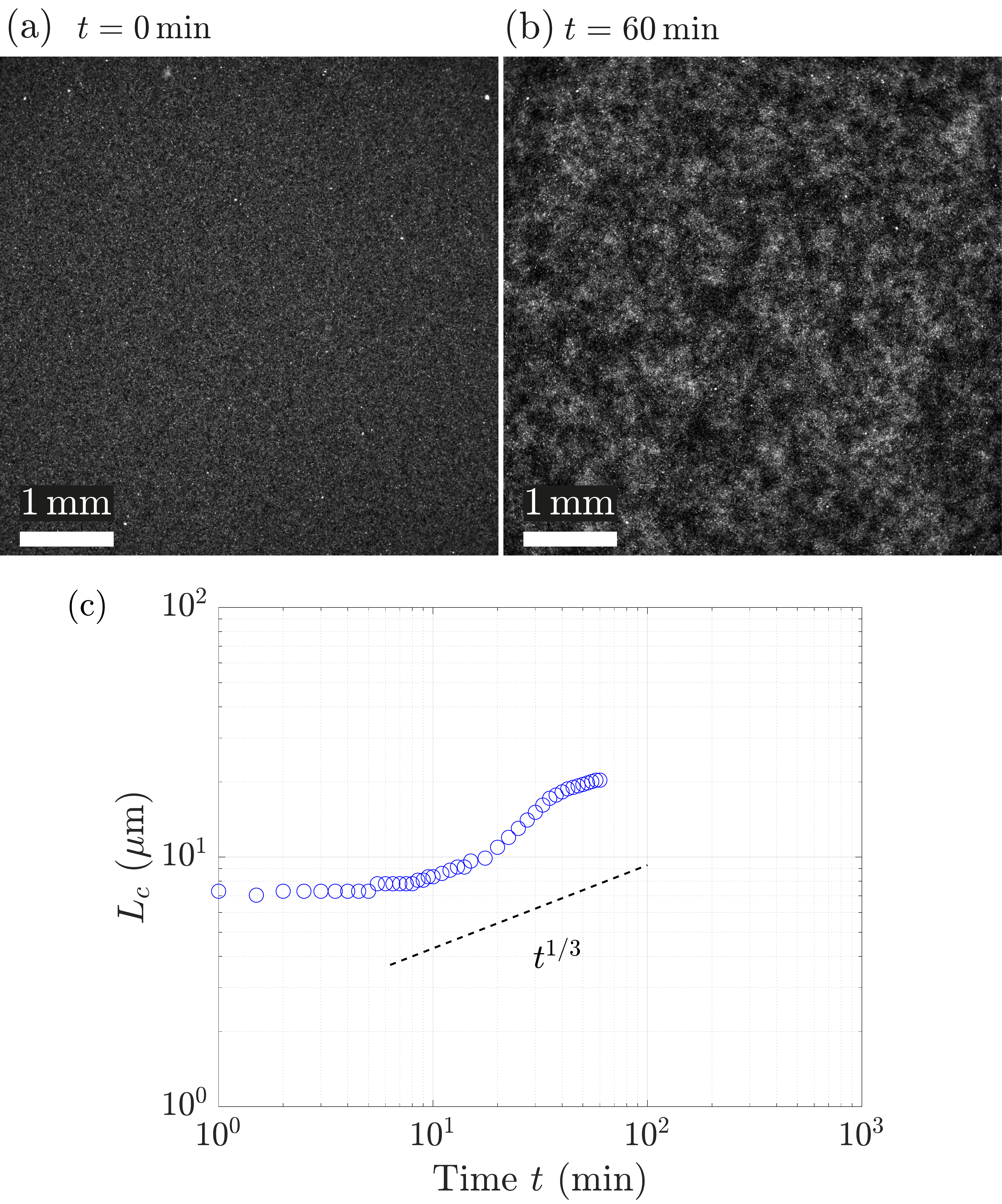}
  \caption{Clustering of beads in a suspension of \emph{E. coli} RP437 bacteria (bead diameter $D_B=\SI{2}{\um}$, surface fraction $\Phi_B=0.30$, $\mathrm{OD}=5$), at $t=0$ (a) and $t=\SI{60}{min}$ (b). (c) Time evolution of the cluster size $L_c(t)$, showing a dynamics similar to the one observed for \emph{B. contaminans}: a plateau at short time followed by an increase at large time, compatible with a $t^{1/3}$ power law.}
  \label{fig:SM_Ecoli}
\end{figure}

To check if the observed bead clustering is specific to the bacterial species \emph{B. contaminans}, we performed experiments with another species, \emph{Escherichia coli} RP 437.
Figure~\ref{fig:SM_Ecoli}(a,b) shows images recorded at the beginning of the experiment and after 1\,h. 
The final image shows a clear clustering of beads, similar to that observed with \emph{B. contaminans}. Figure~\ref{fig:SM_Ecoli}(c) shows the evolution of the cluster size $L_c$ computed using the image processing method as described in Section~\ref{sec:SM_cluster_size}. As for \emph{B. contaminans}, $L_c$ is first constant and then increases in time. The increase is consistent with the power-law $\sim t^{1/3}$ observed with \emph{B. contaminans}.

An important difference between the two bacterial species is the duration of the plateau before the growth regime. The crossover between these two regimes is governed by the time of first encounter of diffusing beads, which is shorter for \emph{B. contaminans} than for \emph{E. coli} under the same conditions. This may be due to a difference in swimming activity between the two species, and led us to select \emph{B. contaminans} rather than \emph{E. coli} for our study.

\section{Effective diffusion of beads in a bacterial bath}
\label{sec:SM_mu}

To determine the effective diffusion coefficient of the beads, we analyze their trajectories when placed in a bacterial bath at $\mathrm{OD}=1$ and 5. A low surface fraction is chosen to minimize their interaction and to prevent the formation of clusters. The bead tracking is performed using Trackmate~\cite{tinevez2017trackmate}, from which we compute the mean square displacement
\begin{equation}
    \langle |\delta \mathbf{r}|^2 \rangle = \langle |{\mathbf{r}}_i(t)-\mathbf{r}_i(0)|^2\rangle,
\end{equation}
where $\mathbf{r}_i(t)$ is the position of the bead $i$ and the brackets $\langle\cdot\rangle$ denote the average over time and beads. For a Brownian particle performing a 2D  random walk of steps of duration $\tau_B$ and length $\ell_B=v_B \tau_B$, the mean square displacement follows a ballistic regime $\langle |\delta \mathbf{r}|^2 \rangle= v_B^2t^2$ for $t<\tau_B$ and a diffusive regime $\langle |\delta \mathbf{r}|^2 \rangle=4\mu_B t$ for $t>\tau_B$.

\begin{figure}[tb]
\centering
  \includegraphics[width=8cm]{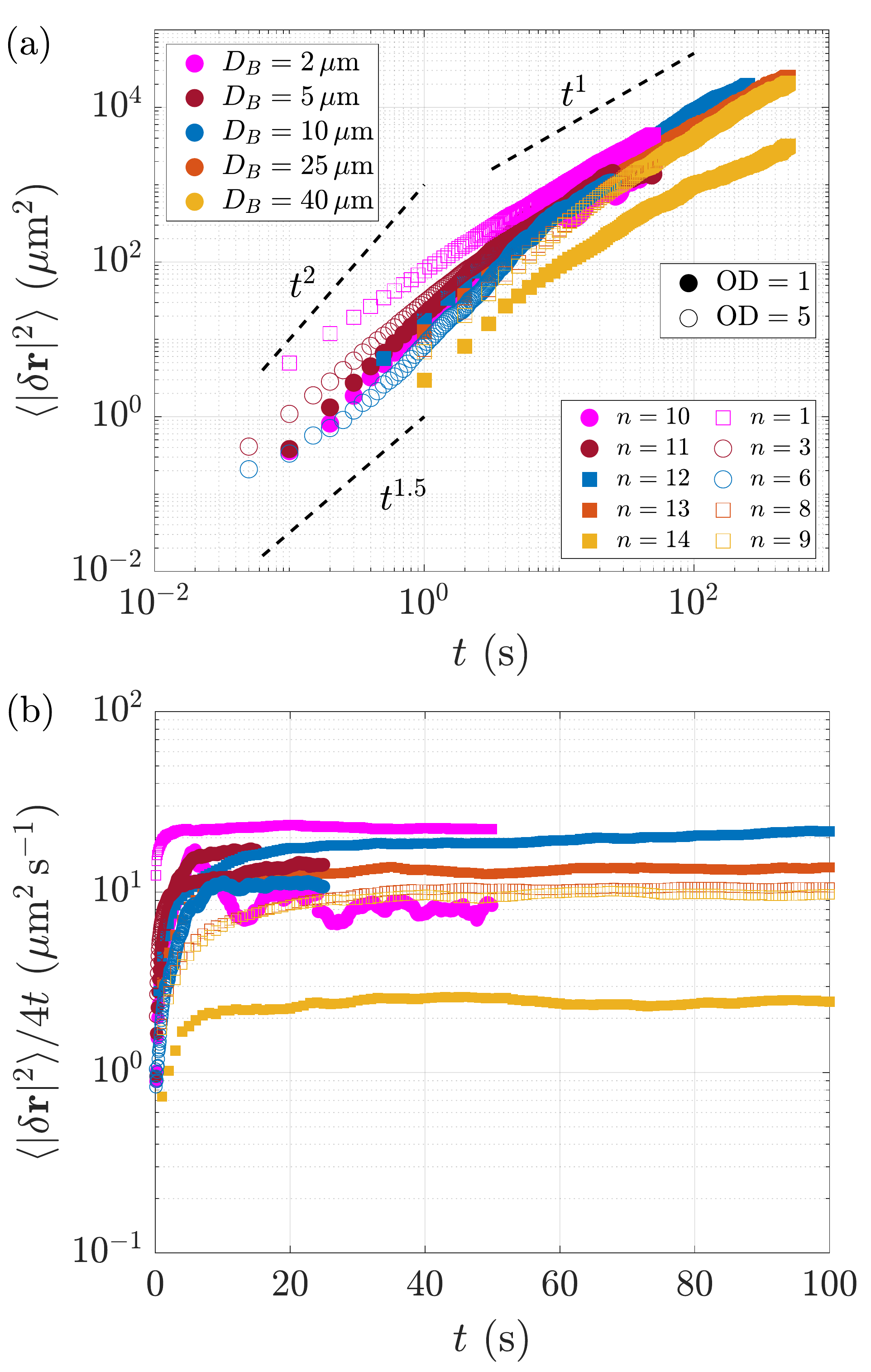}
  \caption{(a) Mean square displacement $\langle |\delta \mathbf{r}|^2 \rangle$ of beads for bacterial concentrations $\mathrm{OD}=1$ (filled symbols) and $\mathrm{OD}=5$ (empty symbols). See Tab.~\ref{tab:SM_diffusion} for experimental conditions. (b) The diffusion coefficient $\mu_B$ is given by $\langle |\delta \mathbf{r}|^2 \rangle/4t$ at large time.}
  \label{fig:SM_MSD}
\end{figure}

Figure~\ref{fig:SM_MSD}(a) shows $\langle |\delta \mathbf{r}|^2 \rangle$ as a function of time for the set of experiments summarized in Tab.~\ref{tab:SM_diffusion}. Our data shows a first regime of slope larger than 1 followed by a diffusive regime, a behavior which is consistent with previous studies~\cite{wu_particle_2000,valeriani2011colloids,patteson2016particle}. To highlight the diffusive regime at large time, we plot the ratio $\langle |\delta \mathbf{r}|^2 \rangle/4t$ in Fig.~\ref{fig:SM_MSD}(b). After 2 to 30 seconds, all curves tend to a constant value, yielding the diffusion coefficient $\mu_B$ reported in Tab.~\ref{tab:SM_diffusion} alongside the experimental conditions. They are plotted as a function of the bead diameter in Fig.~\ref{fig:SM_Diffusion}.
 
\begin{table}[tb]
\centering
\begin{tabular}{cccccccc}
    \toprule
    \toprule
    ~$n$~ & OD & $D_B$ & $C_B$ & $N_{\mathrm{tracks}}$ & MAG & $f_{aq}$ & $\mu_B$ \\
     &  & (\si{\um}) & ~(\si{\g\per\mL})~ &  &  &  (Hz) & (\si{\square\um\per\s}) \\
   \midrule
     1 & 5 & 2 & \num{2e-6} & 23744 & x2.5 & 10  & 23\\    
     2 & 5 & 2 &\num{2e-6} & 32585 & x2.5 & 10  & 21\\
      3 & 5 & 5 &\num{5e-5} & 640  & x10 & 20 & 13\\
      4 & 5 & 5 &\num{5e-5} & 1323  & x10 & 20 & 12\\ 
      5 & 5 & 10 &\num{5e-5} & 741 & x2.5 & 20 & 19\\ 
      6 & 5 & 10 &\num{5e-5} & 328 & x10 & 20 & 11\\     
      7 & 5 & 10 &\num{1e-4} & 103 & x10 & 20 & 23\\
      8 & 5 & 25 &\num{5e-4} & 1079 & x2.5 & 1 & 12\\
     9 & 5 & 40 &\num{1e-3} & 424 & x2.5 & 1 & 10\\     
     \hline
     10 & 1 & 2 &\num{2e-6} & 100 & x10 & 10 & 8.2\\
     11 & 1 & 5 &\num{2e-5} & 612 & x10 & 10 & 16\\
     12 & 1 & 10 &\num{5e-5} & 1224 & x2.5 & 2 & 20\\
     13 & 1 & 5 &\num{2e-4} & 444 & x2.5 & 1 & 13\\
     14 & 1 & 40 &\num{1e-3} & 245 & x2.5 & 1 & 2.5\\
    \bottomrule
    \bottomrule
\end{tabular}
\caption{Summary of the experimental conditions for the determination of the bead diffusivity $\mu_B$: bead diameter $D_B$, bacterial concentration OD, bead concentration $C_B$, number of trajectories $N_{\mathrm{tracks}}$, acquisition frequency $f_{aq}$ and magnification (MAG). The diffusion coefficient $\mu_B$ is computed from the mean square displacement of beads at low concentration, as shown in Fig.~\ref{fig:SM_MSD}.}
\label{tab:SM_diffusion}
\end{table}

\begin{figure}[tb]
\centering
  \includegraphics[width=8cm]{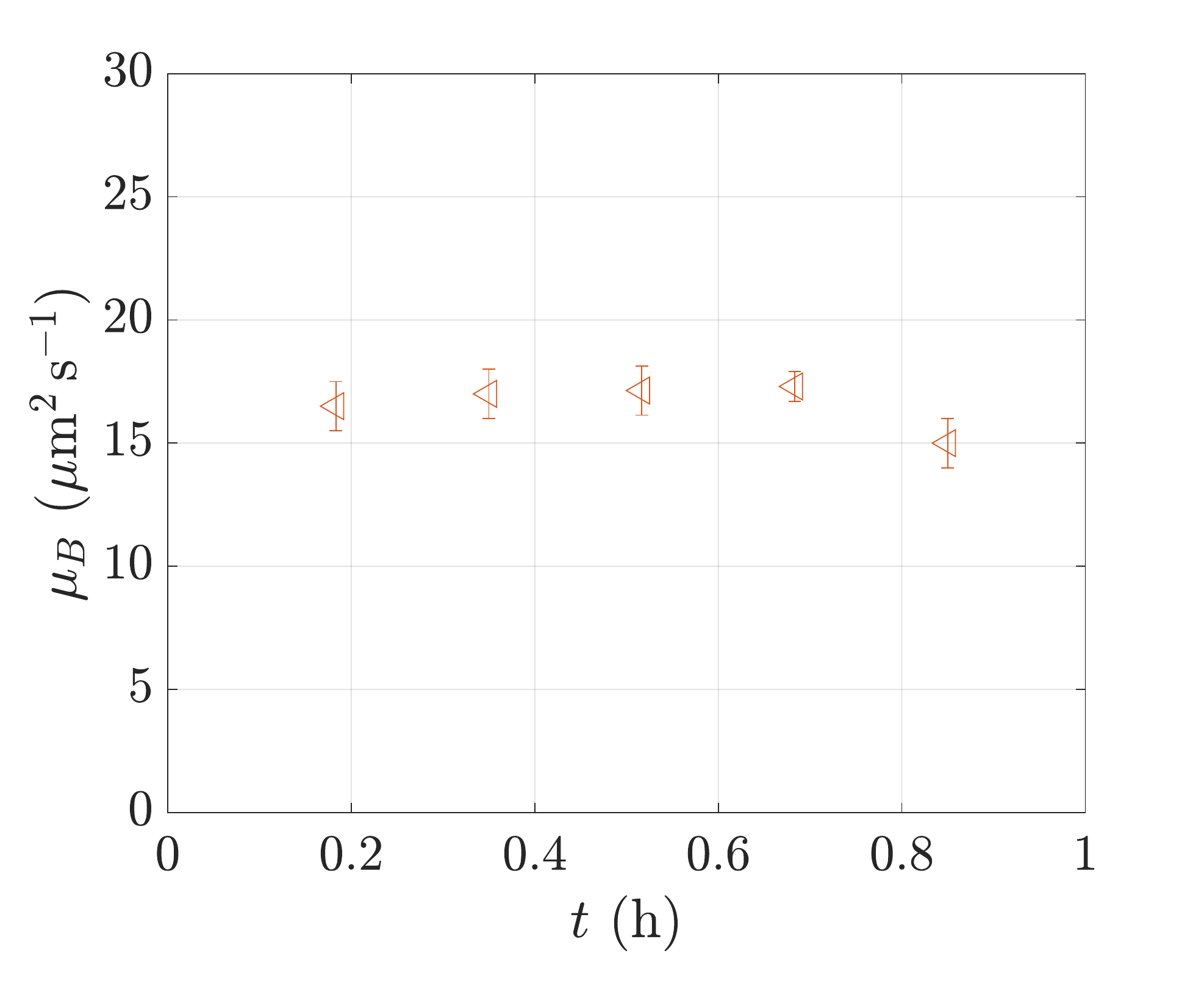}
  \caption{Time evolution of the effective diffusion coefficient of the beads $\mu_B$ during one hour, in the case $\mathrm{OD}=5$, $D_B=\SI{5}{\um}$ and $\Phi_B=0.006$.}
  \label{fig:SM_Diffusion_time}
\end{figure}

Finally, we checked that the bead diffusion coefficient remained constant during the clustering experiments. An experiment was conducted with $\SI{5}{\um}$ beads at a surface fraction $\Phi_B=0.006$ at $\mathrm{OD}=5$. Measurements in Fig.~\ref{fig:SM_Diffusion_time} show a nearly constant $\mu_B$ during 1\,h, which is consistent with the nearly constant bacterial concentration observed in Fig.~\ref{fig:SM_OD}, indicating that the 1-h clustering experiments are performed under stationary bacterial conditions.

\section{Wall-bead friction coefficient $\kappa$}
\label{sec:SM_kappa}

To determine the effective attractive force between beads from our bead velocity measurements, we assume that this force is balanced by the viscous friction. This assumption requires to know the force-velocity relation for a bead moving close to a surface. The bead motion is a combination of rolling and sliding, and depends on several parameters such as the roughness of the beads and substrate.

In order to determine the force-velocity relation, we performed experiments with beads drifting under gravity along the bottom surface of a chamber tilted by a small angle $\theta$ with respect to the horizontal. The chamber is prepared following the same protocol as in Appendix~\ref{sec:SM_bact_prep}, and is filled with the same culture medium as in the clustering experiments but without bacteria. The chamber is mounted on the microscope stage, which is tilted at an angle $\theta$ between \SI{5}{\degree} and \SI{17}{\degree}.

\begin{figure}[tb]
\centering
  \includegraphics[width=8cm]{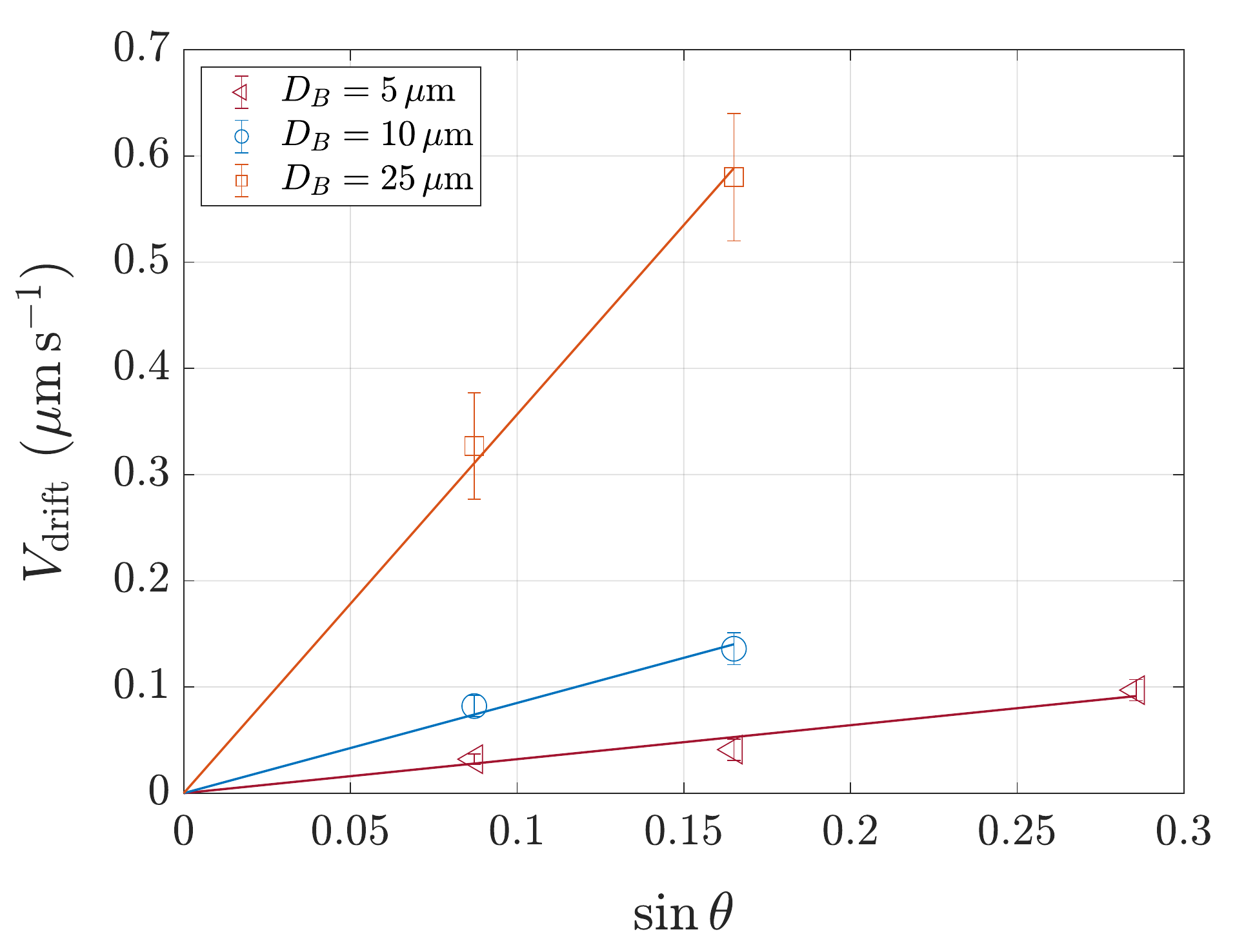}
  \caption{Average drift velocity $V_\mathrm{drift}$ of the beads as a function of the sine of the tilt angle $\theta$, for three bead diameters $D_B=5$, $10$ and \SI{25}{\um}.}
  \label{fig:velocity_vs_alpha}
\end{figure}

The component of the bead weight along the surface is $4/3 \, \pi R_B^3 \Delta \rho g \sin \theta$, with $g=\SI{9.81}{\m\per\square\s}$ the gravitational acceleration, $\eta=\SI{1}{\milli\Pa\s}$ the fluid viscosity and $\Delta \rho=\rho_B-\rho_f$ the density difference between the beads ($\rho_B=\SI{1.06}{\g\per\cubic\cm}$) and the fluid ($\rho_f=\SI{1.007}{\g\per\cubic\cm}$). We write the friction force as $F=\kappa \eta R_B V_\mathrm{drift}$, with $V_\mathrm{drift}$ the drift velocity of the bead, and $\kappa$ a constant accounting for the wall-bead interaction (the standard Stokes force yields $\kappa=6\pi$ far from the walls). The drift velocity is therefore given by:
\begin{equation}
    V_\mathrm{drift} = \frac{4 \pi R_B^2 \Delta \rho g}{3 \kappa \eta} \sin \theta.
    \label{eq:theta}
\end{equation}
The drift velocity was determined from the tracking of dozens of beads for different tilt angles. Figure~\ref{fig:velocity_vs_alpha} shows  $V_\mathrm{drift}$ as a function of $\sin \theta$ for different bead diameters. 
For a given bead diameter, $V_\mathrm{drift}$ grows linearly with $\sin \theta$, as predicted by Eq.~\eqref{eq:theta}. From the measured slope we determine the friction coefficient, and obtain $\kappa=34$, 52 and 77 ($\pm 10\%$) for beads of diameter 5, 10 and \SI{25}{\um}, respectively. These values are used to compute the effective force $F$ defined in Eq.~(\ref{eq:kappa}) and plotted in Fig.~\ref{fig:force}.

\bibliography{Clustering2}

\end{document}